\documentclass[aps,10pt,prb,twocolumn,letterpaper,superscriptaddress]{revtex4}

\usepackage{amsmath}
\usepackage{epsfig}
\usepackage{graphicx, caption}
\usepackage{latexsym}
\usepackage{amssymb}
\usepackage{color}
\usepackage{amsfonts}
\usepackage{bm}
\usepackage{upgreek}
\usepackage[dvipsnames]{xcolor}

\begin{document}
% \title{Glassy to crystalline dopant phase control in electrochemically treated cuprate thin-film structures}
\title{Control of dopant crystallinity in electrochemically treated cuprate thin films}
\author{A.~Frano}
\email{afrano@ucsd.edu}
\affiliation{Department of Physics, University of California, San Diego, California 92093, USA}
\affiliation{Department of Physics, University of California, Berkeley, California 94720, USA}

\author{M.~Bluschke}
\thanks{M.B. and A.F. contributed equally to this work.}
\affiliation{Max Planck Institute for Solid State Research, Heisenbergstr. 1, 70569 Stuttgart, Germany}
\affiliation{Helmholtz-Zentrum Berlin f\"{u}r Materialien und Energie, Wilhelm-Conrad-R\"{o}ntgen-Campus BESSY II, Albert-Einstein-Str. 15, 12489 Berlin, Germany}

\author{Z.~Xu}
\affiliation{Department of Physics, University of California, Berkeley, California 94720, USA}

\author{B.~Frandsen}
\affiliation{Department of Physics, University of California, Berkeley, California 94720, USA}

\author{Y.~Lu}
\affiliation{Max Planck Institute for Solid State Research, Heisenbergstr. 1, 70569 Stuttgart, Germany}

\author{M.~Yi}
\affiliation{Department of Physics, University of California, Berkeley, California 94720, USA}

\author{R.~Marks}
\affiliation{Stanford Synchrotron Radiation Lightsource, SLAC National Accelerator Laboratory, Menlo Park, CA 94025, USA}

\author{A.~Mehta}
\affiliation{Stanford Synchrotron Radiation Lightsource, SLAC National Accelerator Laboratory, Menlo Park, CA 94025, USA}

\author{V.~Borzenets}
\affiliation{Stanford Synchrotron Radiation Lightsource, SLAC National Accelerator Laboratory, Menlo Park, CA 94025, USA}

\author{D.~Meyers}
\affiliation{Condensed Matter Physics and Materials Science Department, Brookhaven National Laboratory, Upton, New York 11973, USA}

\author{M. P. M.~Dean}
\affiliation{Condensed Matter Physics and Materials Science Department, Brookhaven National Laboratory, Upton, New York 11973, USA}

\author{F.~Baiutti}
\affiliation{Max Planck Institute for Solid State Research, Heisenbergstr. 1, 70569 Stuttgart, Germany}

\author{J.~Maier}
\affiliation{Max Planck Institute for Solid State Research, Heisenbergstr. 1, 70569 Stuttgart, Germany}

\author{G.~Kim}
\affiliation{Max Planck Institute for Solid State Research, Heisenbergstr. 1, 70569 Stuttgart, Germany}

\author{G.~Christiani}
\affiliation{Max Planck Institute for Solid State Research, Heisenbergstr. 1, 70569 Stuttgart, Germany}

\author{G.~Logvenov}
\affiliation{Max Planck Institute for Solid State Research, Heisenbergstr. 1, 70569 Stuttgart, Germany}

\author{E.~Benckiser}
\affiliation{Max Planck Institute for Solid State Research, Heisenbergstr. 1, 70569 Stuttgart, Germany}

\author{B.~Keimer}
\affiliation{Max Planck Institute for Solid State Research, Heisenbergstr. 1, 70569 Stuttgart, Germany}

\author{R. J. Birgeneau}
\email{robertjb@berkeley.edu}
\affiliation{Department of Physics, University of California, Berkeley, California 94720, USA}
% \affiliation{Department of Materials Science and Engineering, University of California Berkeley, Berkeley, CA 94720, USA.}

\begin{abstract}
%We have used high-resolution inelastic x-ray scattering (IXS) in combination with Raman scattering to present an experimental survey of the lattice dynamics of bilayer superconducting cuprate compounds. 

We present a methodology based on \textit{ex-situ} (post-growth) electrochemistry to control the oxygen concentration in thin films of the superconducting oxide La$_2$CuO$_{4+y}$ grown epitaxially on substrates of isostructural LaSrAlO$_4$. The superconducting transition temperature, which depends on the oxygen concentration, can be tuned by adjusting the pH level of the base solution used for the electrochemical reaction. As our main finding, we demonstrate that the dopant oxygens can either occupy the interstitial layer in an orientationally disordered state or organize into a crystalline phase via a mechanism in which dopant oxygens are inserted into the substrate, changing the lattice symmetry of both the substrate and the epitaxial film. We discuss this mechanism, and describe the resulting methodology as a platform to be explored in thin films of other transition metal oxides.

\end{abstract}

\maketitle

\section{Introduction}

The design of new materials and control of their properties relies on the development of techniques for fine-tuning internal material parameters. In material systems with strongly correlated electrons\cite{Imada1998} exotic physical properties can be explored by adjusting lattice symmetry, carrier doping, chemical pressure, external pressure and fields. For example, one of the most extensively investigated phase diagrams generated by parameter tuning is that of the superconducting copper oxides\cite{Taillefer2010,Keimer2015}. Introducing charge carriers into an antiferromagnetic Mott insulator renders unconventional superconductivity and a host of other collective quantum phenomena. A modern effort to control both the micro- and macroscopic properties of highly correlated electron systems, while maintaining relatively low disorder levels, has focused on the growth of metastable materials in the form of epitaxial thin films or heterostructures, particularly in the case of transition metal oxides. Application of control parameters such as epitaxial strain, reduced dimensionality, interfacial charge transfer and exchange interactions is beginning to yield a new generation of properties and potential functionalities~\cite{Mannhart2008,Hwang2012,Chakhalian2012, Varignon2018}. 

Nonetheless, control over oxygen stoichiometry and oxygen ordering remains limited in oxide thin films. Oxygen stoichiometry can be manipulated by varying the oxygen partial pressure during growth, but only a limited set of material phases can be synthesized within the range of oxygen pressures compatible with state-of-the-art deposition techniques. \textit{Ex-situ} control of oxygen stoichiometry is commonly achieved by annealing as well as electrochemical procedures. However, due to the extremely small sample volume and significant variations of substrate surface quality, reproducible procedures can be difficult to establish for epitaxial thin films. Ground state magnetic, electronic and ionic orders, as well as electronic and ionic conductivities, are highly susceptible to variations in oxygen stoichiometry and disorder. As such, control of these is essential to the fundamental study of oxide materials and crucial when tailoring material properties for application in a wide range of devices~\cite{Sunarso2017, Hong2018, Varignon2018}. For example, oxygen off-stoichiometry is frequently tuned to optimize electrical and ionic conductivities in perovskite materials used as oxygen-transporting membranes or as elements of solid oxide fuel cells~\cite{Sunarso2017}. Similarly reversible oxygen doping of Ca$_3$Co$_4$O$_9$ films grown on SrTiO$_3$ and LaAlO$_3$ has been shown to enhance significantly the thermoelectric performance of the heterojunction~\cite{Yordanov2017}. In another example, controlled oxygen migration in materials such as VO$_{2-x}$ or Ta$_2$O$_{5-x}$ can produce local metal-insulator transitions which may serve as the basis for resistive random access memory devices~\cite{delValle2017, Hong2018}. In the case of the superconducting copper oxides, oxygen ordering phenomena have been shown to impact both superconductivity as well as the correlated electronic phases with which superconductivity interacts, despite leaving the global stoichiometry of the material unchanged~\cite{Lee2004, Liu2006, Fratini2010, Poccia2011, Bluschke2018}.

In this work we examined films of La$_2$CuO$_4$ grown on LaSrAlO$_4$ (LSAO) substrates as a test platform to compare the effect of electrochemical doping in bulk and in thin films. A series of superconducting transition temperatures with maximum $T_\text{c}$ comparable to those in the bulk were produced by electrochemical doping using various pH levels. For sufficiently elevated pH, dopant oxygens penetrate the LSAO substrate resulting in a modified crystal symmetry. This symmetry change is mimicked by the epitaxial film and results in the crystallization of a dopant oxygen superstructure within the La$_2$CuO$_{4+y}$ interstitial layers  (Fig.~\ref{f:f1}a,b). 
% This symmetry change is mimicked by the epitaxial film and results in a glassy to crystalline transition of the interstitial oxygens (Fig.~\ref{f:f1}a,b). 

\section{Methods}

An atomic layer-by-layer, ozone-assisted molecular beam epitaxy system was used to grow high quality films of undoped La$_2$CuO$_4$ with a thickness of 30~nm. The growth was performed at a temperature of 630--650$^{\circ}$C in a 1.5$\times$10$^{-5}$ mbar ozone atmosphere and the samples subsequently cooled at the same pressure. Since the synthesis was performed in oxidation (ozone) conditions, the La$_2$CuO$_{4}$ films were vacuum annealed at 200$^{\circ}$C following growth to remove any interstitial oxygens. The fluxes of the effusion cells were calibrated with a quartz crystal monitor before growth. Reflection high-energy electron diffraction was used for real-time monitoring of the synthesis, enabling atomic layer control. The substrates were commercially available (001)-oriented LaSrAlO$_4$ (LSAO) substrates with one side polished.

An electrochemical reaction was used to obtain different doping levels in the samples. Controlled amounts of solid NaOH were dissolved into distilled water generating alkaline solutions of varying pH. The resulting pH levels were measured using conventional pH indicator strips. In the solution the sample was enveloped in a platinum mesh and a potential difference between a platinum counter electrode and the film+substrate was used to produce an electric field which drives apart the Na$^+$ and OH$^-$ ions. A schematic of the electrochemical doping setup and reaction is shown in Fig.~\ref{f:f1}c-d. Half reactions occur~\cite{Grenier1992, Chou1992} at the working and counter electrodes, respectively: $\mathrm{La}_2\mathrm{Cu}\mathrm{O}_4+2y\mathrm{OH}^-\to \mathrm{La}_2\mathrm{Cu}\mathrm{O}_{4+y}+y\mathrm{H}_2\mathrm{O}+2ye^-$ and $2y\mathrm{H}_2\mathrm{O}+2ye^-\to 2y \mathrm{OH}^-+y\mathrm{H}_2(g)$. The net electrochemical reaction pertaining to the film is $\mathrm{La}_2\mathrm{Cu}\mathrm{O}_4+y\mathrm{H}_2\mathrm{O}\to \mathrm{La}_2\mathrm{Cu}\mathrm{O}_{4+y}+y\mathrm{H}_2$. %Given a sufficient oxidizing power, a similar reaction occurs in the substrate as well.

 All electrochemical processes were performed in a potentiostatic mode at room temperature for approximately 18 hours. The voltage, which was applied using a commercial power supply and monitored by a reference electrode, was 0.6~V. This value is below the threshold for an oxygen evolution reaction ($2\mathrm{OH}^{-}\to\mathrm{H}_2\mathrm{O}+ \frac{1}{2}\mathrm{O}_2+2\mathrm{e}^-$) to happen at the working electrode~\cite{Grenier1992, Chou1992}. The resulting currents upon turning on the voltage were $\sim$ 0.6~mA but would drop about an order of magnitude after 12 hours.

A magnetometric characterization of the resulting superconducting transitions is presented in Fig.~\ref{f:f1}e. The magnetic susceptibility were measured using a vibrating sample magnetometer -- superconducting quantum interference device (SQUID). Because of the substrate contribution to the signal, absolute superconducting volume fractions are impossible to obtain.

 % The process of electrochemically doping single crystals of bulk La$_2$CuO$_4$ in order to yield superconductivity is known to be time consuming. Depending on the crystal size, it can take up to several months in the solution~\cite{Radaelli1993}. In contrast, doping of nanoscale films such as the 30 nm films studied here can be accomplished in less than 24 hours.

The X-ray scattering experiments were performed at beam line 10-2 of the Stanford Synchrotron Radiation Lightsource at the Stanford Linear Accelerator Center in Menlo Park, California, USA. X-rays with photon energy 9~keV were used to access a large Ewald sphere. The beam spot size was $\sim$300$\times$300 microns. The scattered intensity was measured with an avalanche photodiode and with a solid state detector. The sample was cooled using a Displex cryostat. After electrochemical oxidation, all samples were stored under ambient conditions (300~K) for 1-2 days before quench cooling to cryogenic temperatures to perform the reciprocal space maps. Temperature dependences were performed upon warming after quench cooling from 300~K.

\section{Results and Discussion}

\begin{figure*}[htb]
	\captionsetup{justification=raggedright,width=1.8\columnwidth}
	\includegraphics[width=1.8\columnwidth]{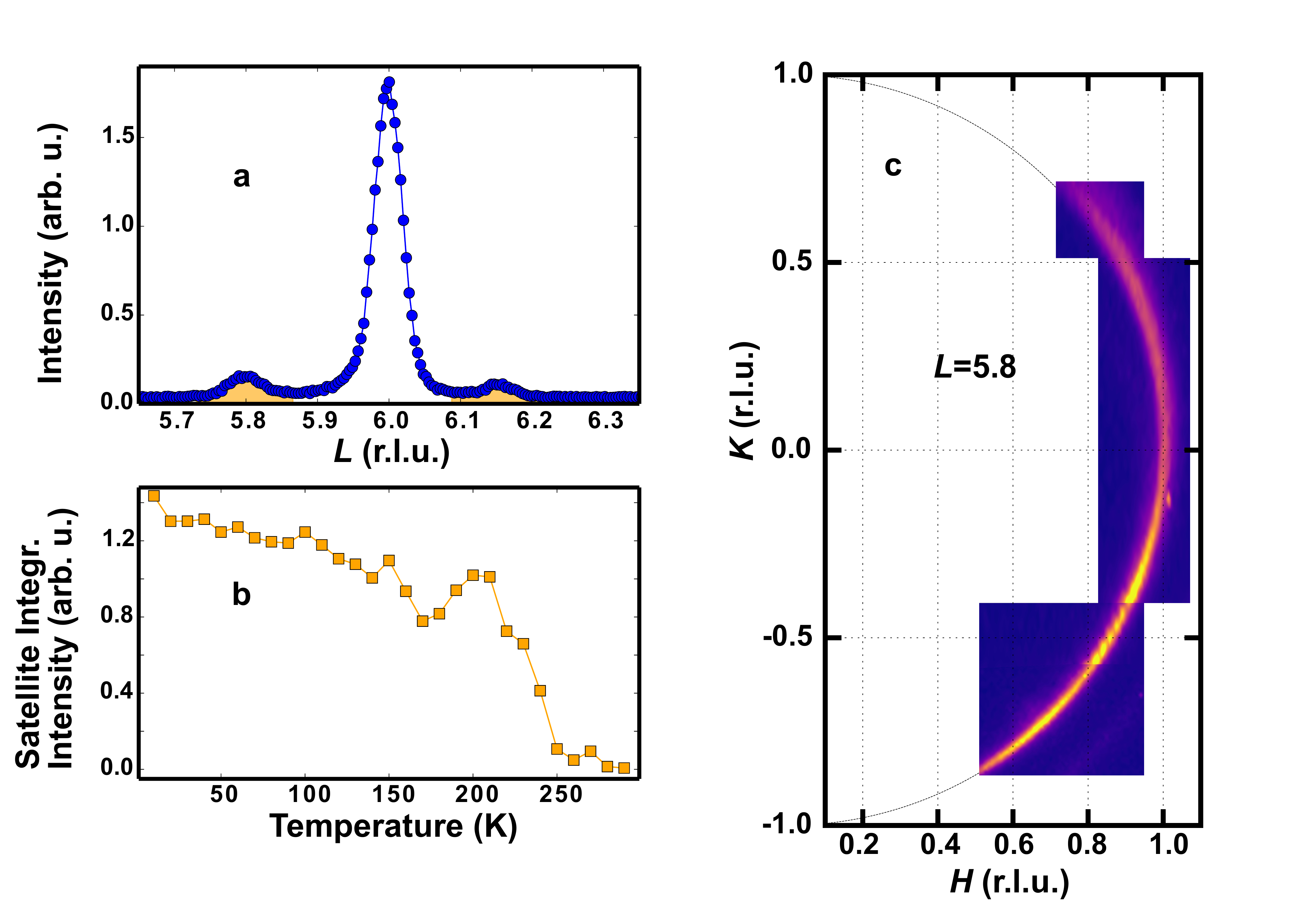}
	\caption{For a sample doped in a pH=10 solution: (a) X-ray diffraction $L$-scan over the (0,1,6) reflex of La$_2$CuO$_{4+y}$ displaying two satellite peaks which correspond to staging with period $\sim6$. (b) Integrated intensity of the satellite peaks as a function of temperature, where the shaded regions in panel (a) denote the integration ranges. (c) Reciprocal space map of the experimentally accessible (H,K)-plane for $L$=5.8 taken at $T$=10~K. The dotted circle indicates in-plane Q-vectors of equal length.}
	\label{f:f2}
\end{figure*}

We begin by discussing the effect of electrochemically doping the film+substrate system with alkaline solutions of relatively low base concentration, namely pH~$\lesssim$~11. In this case, only the La$_2$CuO$_{4}$ film is oxidized. We readily observe the segregation of dopant oxygens into evenly spaced interstitial planes along the $c$-axis. This out-of-plane oxygen ordering known as `staging' results in an antiphase boundary of the CuO$_6$ octahedral tilt pattern at the interstitial layers (schematically depicted in Fig.~\ref{f:f1}a) and is well documented in the case of bulk La$_2$CuO$_{4+y}$~\cite{Radaelli1993, Wells1996, Wells1997, Blakeslee1998, Lee1999, Wochner1996, Xiong1996}, but has not yet been reported in films of the same material~\cite{Locquet1993, Arrouy1996, Wang2004}. Fig.~\ref{f:f2}a shows a representative X-ray diffraction $L$-scan around the (0,1,6) reflection associated with the unmodulated tilt pattern recorded at $T=10$~K for a sample processed in a solution of pH 10 (throughout this report reciprocal lattice coordinates are labelled in the orthorhombic $Bmab$ setting of the film). Two satellites are clearly visible with an average separation from the central peak corresponding to a staging period of approximately 5.7 lattice parameters, which we refer to as stage 6 (equivalent to 1 oxygen interstitial plane every 3 unit cells, see Fig.~\ref{f:f1}a). 
We attribute the incommensurability of the staging peak to the presence of faults in the c-axis periodicity of the staging pattern. %Note that the system shows a miscibility gap precluding stage 5 ordering~\cite{Wells1996}.
The temperature dependence of the area under the satellites is shown in Fig.~\ref{f:f2}b. As in the bulk, the onset of stage 6 order is sharp around $T=250$~K~\cite{Wells1996, Wochner1996, Xiong1996}. We emphasize at this point that the information about interstitial oxygen ordering presented in this study is obtained indirectly by measuring the staging diffraction peaks which are the result of scattering from modulations of the CuO$_6$ octahedral tilt pattern. These modulations are induced by the presence of the interstitial oxygens and are therefore sensitive to their $c$-axis periodicity and the symmetry of their in-plane ordering.

A striking difference with respect to previous studies on the bulk system is seen in the in-plane spread of the staging satellite reflections. Fig.~\ref{f:f2}c shows an X-ray diffraction reciprocal space map of the ($H$,$K$,$L=5.8$) plane at $T=10$~K. Whereas in bulk La$_2$CuO$_{4+y}$ the staging reflections are peaked at in-plane coordinates ($H$,$K$)=(1,0) and (0,1) consistent with a checkerboard pattern, the staging reflection shown in Fig.~\ref{f:f2}c is peaked along $L$ but spread into a ring-like distribution of intensity within the $H$-$K$ plane  with a constant in-plane momentum $|($H$,$K$)|=1$, i.e. consistent with a commensurate one-unit-cell periodicity with random in-plane orientation.  This novel diffraction ring indicates a reduction of orientational order whose origin is not clear. By considering the ring-like staging peaks alone, it would be natural to conclude that the electrochemical treatment has caused the La$_2$CuO$_{4+y}$ films to disintegrate into crystallites or domains that undergo random rotations about the $c$-axis. However, we point out that the La$_2$CuO$_{4+y}$ non-staging Bragg peaks - i.e., integer $L$-values with ($H$,$K$)=(1,0) and (0,1) - remain sharp in both in-plane directions, indicating that the crystallinity of the films is preserved. These apparently contradicting observations seem to simultaneously indicate that the Cu-O-Cu bond directions are randomly oriented within the $H$-$K$ plane, and that they obey the $C4$ symmetry of the twinned lattice. Intriguingly however, rather than a phase separation scenario, the comparable widths along $L$ of the two reflections is consistent with a scenario in which both signals arise from a single structure. Since the ring of intensity in the $H$-$K$ plane is only present at $L$-values that correspond to the staging order, we conclude that any disorder phenomenon evidenced by this diffraction pattern must be associated with the staging of the electrochemically doped ions. Furthermore, the staging diffraction ring is sharp in the radial direction, suggesting the pattern originates from features within the interstitial layers that inherit the periodicity and extended correlation length of the preexisting orthorhombic tilt pattern.  The broad angular distribution of the intensity reveals an orientational disorder which could possess an interesting resemblance to two-dimensional glassy networks\cite{Brock1989, Castro2005I,Zaluzhnyy2017}. %This next paragraph is only on the ARXIV version:
In any event, the precise microscopic organization of interstitial oxygens and tilted octahedra which gives rise to the reduced orientational order observed in Fig.~\ref{f:f2}c remains an open question of this work. Although we are not able to unambiguously identify a microscopic model associated with the reduced orientational order, we wish to draw the reader's attention to a recent report \cite{Kang2019} in which charge ordering in electron doped T'-(Nd,Pr)$_2$CuO$_4$ thin films results in a planar diffraction ring while the structural crystallinity of the films is preserved. The case of electron doped T'-(Nd,Pr)$_2$CuO$_4$ thin films presents an intriguing analog to our results and raises the question as to whether the reduced orientational order observed in electrochemically doped La$_2$CuO$_{4+y}$ might be connected to a related Fermi-surface-dependent charge-ordering phenomenon. We now turn to discuss the effect of using an electrochemical solution with a higher pH level.

\begin{figure*}[htb]
	\captionsetup{justification=raggedright,width=1.8\columnwidth}
	\includegraphics[width=1.8\columnwidth]{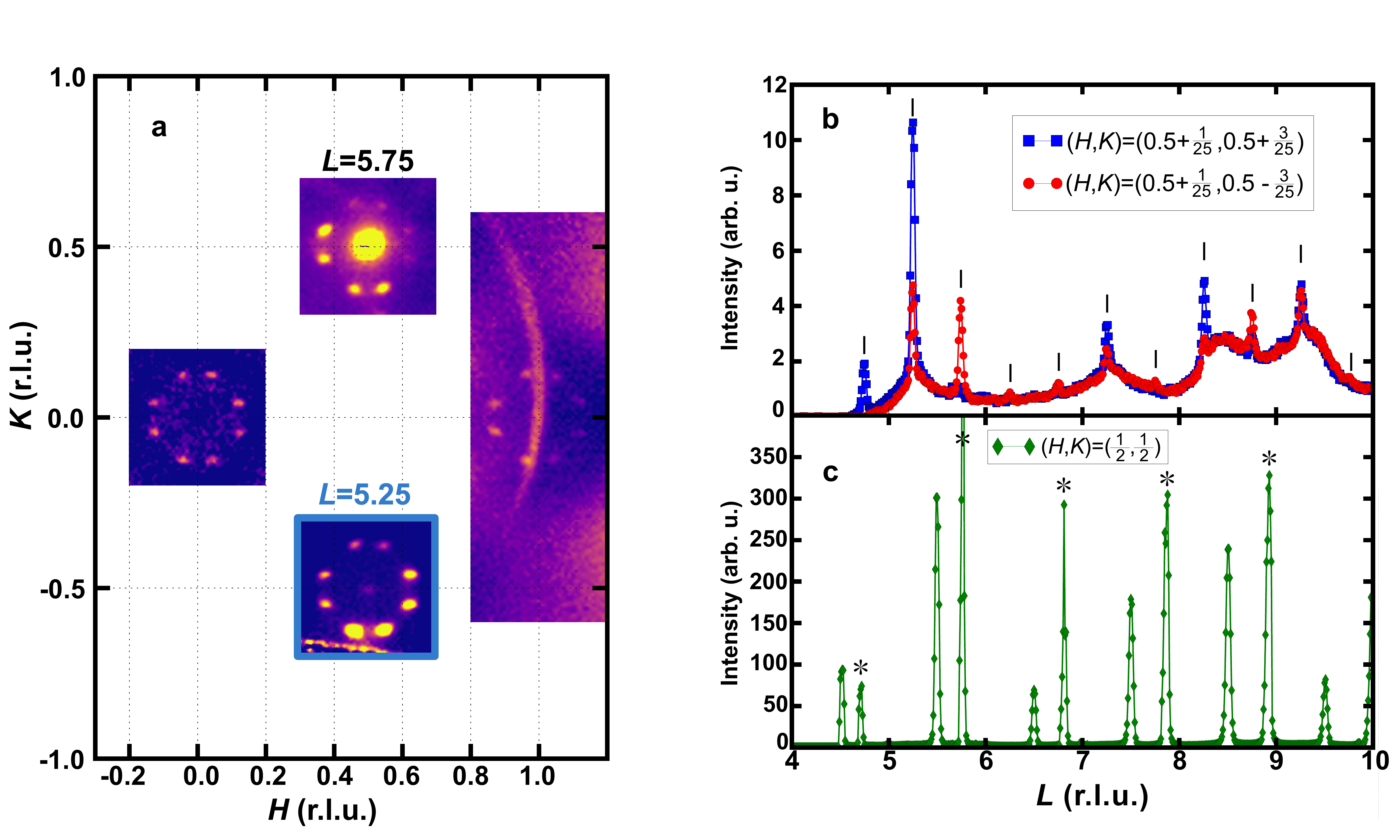}
	\caption{For a sample doped in a pH=14 solution: (a) X-ray diffraction reciprocal space map of the ($H$,$K$)-plane at $L$=5.75 and $L$=5.25 (outline in blue) measured at $T$=10~K. A cluster of 8 peaks around commensurate Bragg positions as well as a remnant of the diffraction ring near (0,1,5.75) can be observed. (b) X-ray diffraction scans along $L$ at in-plane ($H$,$K$) values corresponding to two of the oxygen superstructure peaks; blue squares at ($\frac{27}{50}$,$\frac{31}{50}$) and red circles at ($\frac{27}{50}$,$\frac{19}{50}$). Vertical lines are drawn at every stage-4 position. (c) Scans along $L$ at ($H$,$K$)=($\frac{1}{2}$,$\frac{1}{2}$) where the peaks identified by a $\ast$ symbol occur at half-integer values of $L$ in reciprocal lattice units of the substrate. }
	\label{f:f3}
\end{figure*}

\begin{figure*}[htb]
	\captionsetup{justification=raggedright,width=1.8\columnwidth}
	\includegraphics[width=1.8\columnwidth]{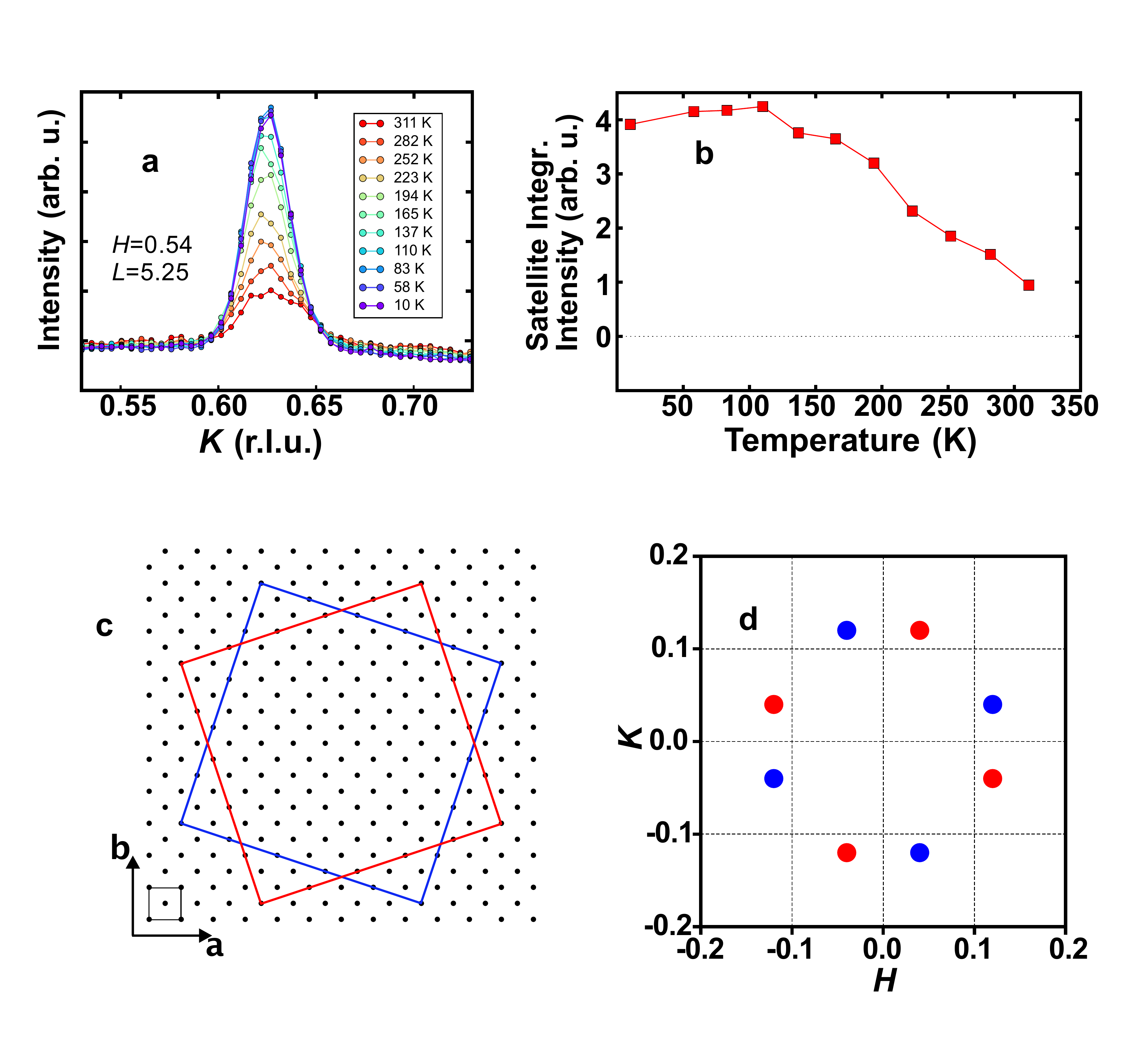}
	\caption{(a) Reciprocal space scans of one representative stage 4 in-plane superstructure peak for different temperatures, as labelled. (b) The integrated intensity of the stage 4 peak plotted as a function of temperature. (c) A representation in real space of the in-plane unit cells associated with the two superstructure domains. The black dots represent the planar Cu ions. A black square denotes the size of the orthorhombic in-plane unit cell. For simplicity, the orthorhombic distortion is not shown.  (d) In reciprocal space, each of the domains create a 4-peak multiplet (colour coded according to the left panel) as observed in the data.}
	\label{f:f4}
\end{figure*}

Upon increasing the pH level of the electrolyte solution to 12.5 and higher, the dopant oxygen ions were observed to crystallize into a well ordered structure within the interstitial layer. Fig.~\ref{f:f3}a shows reciprocal space maps of the $H$-$K$ plane for pH 14 at $L$-values 5.25 and 5.75 corresponding to stage 4 ordering. A cluster of 8 sharp peaks can be observed in the vicinity of various high-symmetry in-plane reciprocal space positions such as ($H$, $K$)~=~(0,0), (0.5,0.5), (0,1), with splitting values of ($H\pm\frac{1}{25}$,$K\pm\frac{3}{25}$) and ($H\pm\frac{3}{25}$,$K\pm\frac{1}{25}$). These staging peaks are as sharp as the main film Bragg peaks in all three reciprocal space directions, corresponding to a correlation length of order 10 nm. The correlation lengths quoted in this report are defined as $\xi_{a,b,c} = \frac{a, b, c }{\pi \times FWHM}$ where FWHM is the full width at half maximum extracted from Lorentzian fits to the data. Due to technical limitations the temperature dependence of this oxygen order (shown in Fig.~\ref{f:f4}a,b) could only be measured up to 311~K, however an extrapolation of the measured trend indicates an onset temperature of approximately $350 \pm 20$ K. This  is consistent with the in-plane ordering of interstitial oxygens reported in Refs.~\citenum{Lee2004} and~\citenum{Fratini2010}, where the onset temperatures were found to be $\sim$330 K and $\sim$350 K respectively. We note however, that the size of the in-plane superstructure observed in our films is doubled with respect to the ordering pattern reported by Refs.~\citenum{Lee2004, Fratini2010, Poccia2011}. Interestingly, a similar superstructure pattern is common in `245' iron selenide superconductors with $\sqrt{5}\times\sqrt{5}$ vacancy order\cite{Zavalij2011, Wang2016}. Just as in the iron selenides, there are two possible rotational domains. These are denoted and colour coded in Fig.~\ref{f:f4}c, along with their layout in reciprocal space in Fig.~\ref{f:f4}d. Fig.~\ref{f:f3}b compares the $L$-dependence of each of the two domains. Each domain has a slightly different $L$-dependent structure factor but the peaks are centered at commensurate values of $L=n\pm\frac{1}{4}$ ($n$ is an integer). In other words, the in-plane superstructure has a well-defined stage 4 order in the out-of-plane direction.

We now discuss how dopant oxygens crystallize into in-plane superstructures in epitaxial La$_2$CuO$_{4+y}$. Although in-plane ordering of oxygen interstitials has been observed in bulk La$_2$CuO$_{4+y}$~\cite{Lee2004, Fratini2010, Poccia2011}, the same has not been reported in epitaxially grown thin films. This may result from the intrinsic disorder in thin film materials which in part originates at the substrate-film interface where lattice mismatch and strain accommodation lead to the formation of structural defects. On the other hand, the substrate-film interface can serve as a template for desired structural distortions in the epitaxial film. In fact, with a sufficient oxidation power we found that it was possible to oxidize the LSAO substrate reducing its lattice symmetry from tetragonal to orthorhombic or even lower symmetry. Due to their epitaxial relationship both substrate and film adopt similar distortion patterns. The following correlation was confirmed in all films studied: whenever the pH was high enough to generate substrate-based structural distortions of the La$_2$CuO$_{4+y}$ crystal structure, an in-plane oxygen superstructure was also observed. The modification of the substrate lattice symmetry is evidenced by the appearance of a series of otherwise forbidden reflections of the substrate, including (1, 0, 6)-type orthorhombic diffraction peaks which are present in the $Bmab$ structure of La$_2$CuO$_4$, but are not present in the tetragonal structure of the undoped LSAO substrate. A further set of reflections, forbidden in both undoped LSAO and undoped La$_2$CuO$_4$, are observed at half-integer reciprocal space coordinates ($\frac{n_h}{2}$,$\frac{n_k}{2}$,$\frac{n_l}{2}$) in units of both the film and the substrate. These are seen in Fig.~\ref{f:f3}c, which shows a reciprocal space scan along the ($\frac{1}{2}$,$\frac{1}{2}$,$L$) direction, demonstrating sharp diffraction peaks at half integer $L$. These peaks are temperature independent up to 350~K implying that a stable lattice distortion was produced \textit{ex-situ} by forcibly intercalating oxygen ions into the substrate. While the in-plane correlation length of the half-order reflections is on the order of 100 nm in the substrate, it is only 10 nm in the film indicating that the structural distortion originates in the substrate and is adopted by the film due to their epitaxial relationship. Furthermore, the appearance of the film staging peaks (Fig.~\ref{f:f3}a) clustered around in-plane reciprocal space positions ($H$,~$K$)~=~($\pm$0.5,~$\pm$0.5) as well as (0,~$\pm$1) and ($\pm$1,~0) indicates that the in-plane oxygen superstructure is associated with a periodic modulation of the substrate-assisted distortion. This distortion provides a structural template which helps to stabilize the oxygen superstructure, whose large correlation length arises naturally from the extended substrate-film interface.

Before being doped, our substrate LSAO has a tetragonal lattice ($a_{LSAO}^{tet}=3.757$~{\AA}, $c_{LSAO}^{tet}~=~12.636$~{\AA}), with spacegroup $I4/mmm$, i.e. no octahedral tilts. When doped, LSAO has a distorted lattice in the plane, with average in-plane lattice parameter $a_{LSAO+\delta}^{ortho}=5.388$~{\AA} (corresponding to $a_{LSAO+\delta}^{tet}=3.812$~{\AA}) and out-of-plane $c_{LSAO+\delta}^{ortho} = 12.637$~{\AA}, i.e. the presence of the additional oxygens in the LSAO structure results in a dramatic $1.7\%$ expansion of the in-plane lattice constant. We speculate that the high density of interstitial oxygens drives the structural distortion, which in turn expands the in-plane lattice constant allowing more room for the intercalated oxygen ions. This new structural distortion in both the substrate and film may be analogous to the monoclinic distortion observed in heavily oxygen doped La$_2$NiO$_{4.25}$ (La$_8$Ni$_4$O$_{17}$)~\cite{Demourgues1993}. Since the dopant oxygen ions do not penetrate into the entire bulk of the LSAO substrate, the lattice parameters after electrochemical oxidation have been extracted from the doping-induced half-order structural reflections. The half-order reflections in the LSAO substrate have an out-of-plane correlation length of 44 nm, evidencing that the dopant oxygens diffused at least this far (though possibly further) into the substrate. Since the flexible oxidation state of the Cu ions in La$_2$CuO$_{4}$ facilitates incorporation of oxygen ions from the NaOH solution~\cite{Grenier1992}, it may be that the electrochemical oxidation of the LSAO substrate is only possible via the La$_2$CuO$_{4}$ electrode. 

Due to the stable closed-shell electronic configuration of all cations in the LSAO substrate (compared to the flexible valence of Cu in La$_2$CuO$_{4}$), the mechanism by which oxidation of the LSAO substrate occurs is not clear. With a DC voltage of 0.6 V and a pH~$\gtrsim$~12.5, it was possible to force oxygen into the LSAO substrate via the conducting La$_2$CuO$_{4}$ film. In order to compensate the charges induced by interstitial oxygens (O$_i$), it is suspected that an oxidation of the regular oxygens (O$_{reg}$) of the 214-structure occurs, $O_{reg}^{2-} + O_{i} \rightarrow O_{reg}^{-} + O_{i}^{-}$, similar to that recently reported in the context of transition-metal-oxide cathodes for Li-ion batteries~\cite{Luo2016}. Additionally, oxides in contact with water are able to take up hydroxide ions by either filling oxygen vacancies or by occupying interstitial sites with charge compensation achieved by uptake of the residual proton. The diffusion coefficient and size requirements of hydroxide ions are typically similar to those of oxygen and their experimental distinction is difficult, in particular in thin films. However, previous studies on electrochemically treated La$_2$CuO$_{4}$ crystals found no indications of OH$^-$ intercalation~\cite{Grenier1992}. %After the electrochemical process, films of La$_2$CuO$_{4+y}$ show clear superconducting transitions. Fig.~\ref{f:f1}c shows magnetization-versus-temperature curves of various samples doped in alkaline solutions with the indicated pH levels. The onset temperature of superconductivity increases with the pH level, suggesting that the doping concentration increases until it reaches optimal levels.
\section{Summary}
In conclusion, we have identified a mechanism which generates unprecedented long-range three-dimensional oxygen order in an oxide film using electrochemical techniques and exploiting the epitaxial connection with the substrate. In our case study of La$_2$CuO$_{4+y}$ we readily achieve both high superconducting transition temperatures and staging behavior of interstitial oxygens consistent with previous studies of bulk La$_2$CuO$_{4+y}$~\cite{Radaelli1993, Wells1996, Wells1997, Blakeslee1998, Lee1999, Wochner1996, Xiong1996}. For moderate oxidation power (pH~$\lesssim$~11) the dopant oxygens form two-dimensional layers with regular stacking along the $c$-axis but are orientationally disordered within the $a$-$b$ plane. For stronger oxidation power (pH~$\gtrsim$~12.5) the substrate can be oxidized resulting in an altered crystal symmetry. The resulting structural template is incorporated by the film and allows the dopant oxygens to crystallize into oxygen ordering superstructures within the interstitial plane. In bulk La$_2$CuO$_{4+y}$ scale-invariant fractal distribution of oxygen order has been shown to enhance superconductivity~\cite{Fratini2010}. Accordingly, thin film La$_2$CuO$_{4+y}$ with tunable oxygen order in the interstitial planes may represent an interesting model system for future studies of order-disorder transitions in two dimensions, and the connection between fractal oxygen ordering and superconductivity. More generally, tunability of oxygen ordering phenomena in oxide thin films and heterostructures is an important step towards harnessing the correlated electronic states with which they interact.

\vspace{10 mm}

\section{Appendix A: Crystal structure of bulk La$_2$CuO$_{4+y}$}

\begin{figure*}[htb]
	\includegraphics[width=1.8\columnwidth]{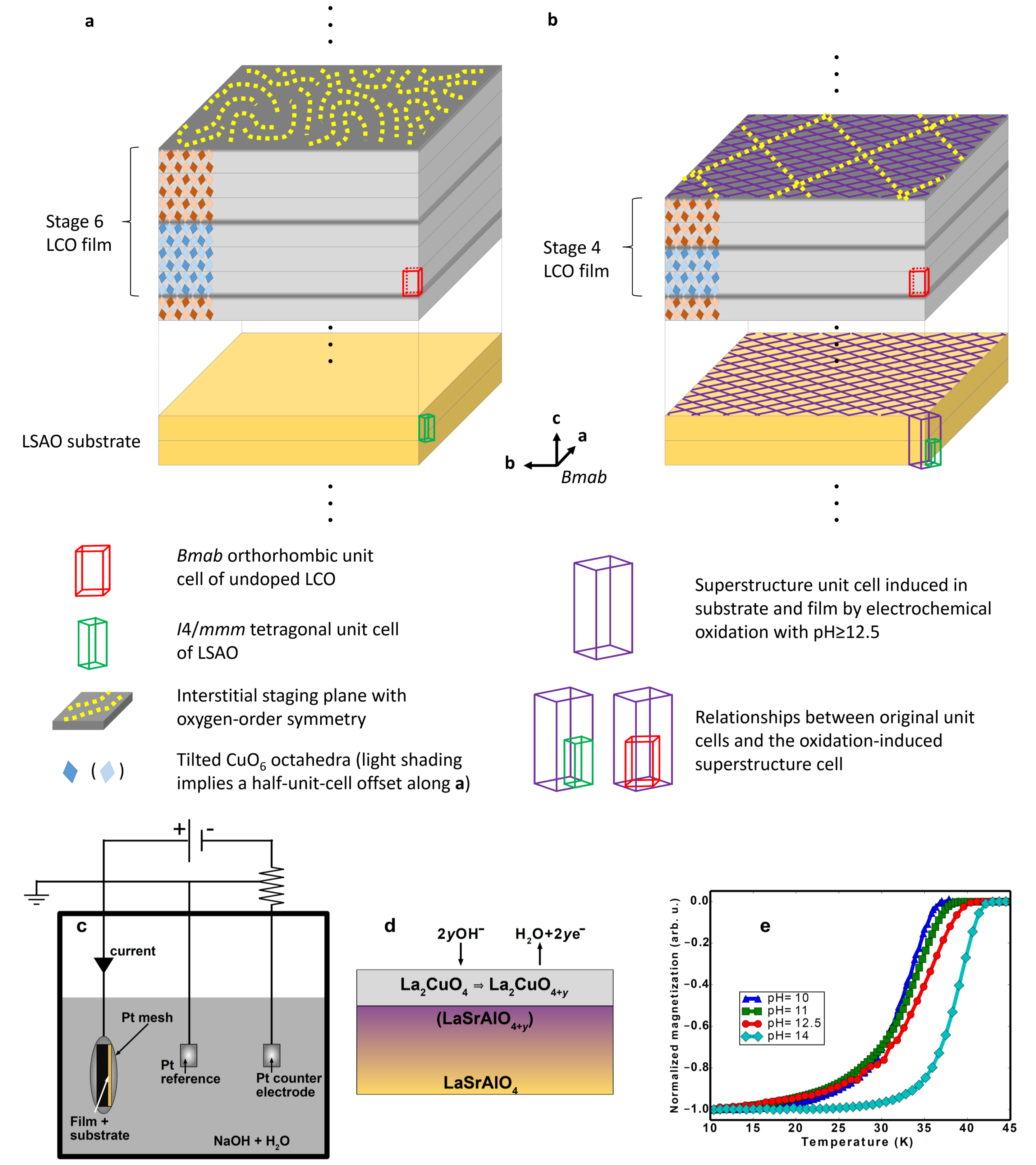}
	\captionsetup{justification=raggedright,width=1.8\columnwidth}
	\caption{(a-b) Schematic representation of the structural unit cells and oxygen ordering patterns in the  La$_2$CuO$_4$/LaSrAlO$_4$ film-substrate system. (a) For electrolyte solutions with pH~$\lesssim$~11 dopant oxygens penetrate the film alone, separating along the c-axis into evenly spaced two-dimensional layers of interstitial oxygens. Within these layers the dopant oxygens are orientationally disordered (depicted schematically by dotted yellow lines). (b) For pH~$\gtrsim$~12.5 dopant oxygens penetrate the substrate effecting a structural distortion (dark purple superstructure) which is incorporated into the film promoting crystallization of the dopant oxygens within the interstitial layers. Dotted yellow lines indicate the dopant-oxygen induced superstructure cell. (c) Schematic of the setup used to electrochemically intercalate oxygen into films of La$_2$CuO$_4$ grown on LSAO. (d) At the working electrode side to which OH$^-$ ions are attracted, the film and the substrate are oxidized. During the short duration of our treatment, the oxygen ions likely penetrate through the film into the substrate. (e) Magnetization normalized to the maximum diamagnetic signal as a function of temperature for samples electrochemically doped in an alkaline solution (H$_2$O+NaOH) with the indicated pH levels.}
	\label{f:f1}
\end{figure*}

Above room temperature the crystal structure of the undoped compound La$_2$CuO$_4$ undergoes a transition from tetragonal to orthorhombic. Below the transition temperature, the CuO$_6$ octahedra of the `perovskite layer' adopt a tilt pattern described by alternating clockwise and counterclockwise rotations about the [100] axis, giving rise to a series of diffraction peaks (called `tilt peaks') with selection rules (0,$K$=odd,$L$=even). These tilts amount to a buckling of the CuO$_2$ planes which helps to accommodate the chemical strain imposed by the rock-salt LaO layer~\cite{Mehta1994}. In the doped compound, La$_2$CuO$_{4+y}$, the dopant oxygens intercalate into the LaO layers, consequently repelling nearby apical oxygen ions of the undoped La$_2$CuO$_{4}$ structure, and producing reversals of the octahedral tilt directions. The kinetics of the dopant oxygens at room temperature allow them to anneal into highly ordered states, characterized by $c$-axis segregation of the dopant oxygens into interstitial planes (whereas the limited mobility of strontium or barium ions in La$_{2-x}$Sr$_{x}$CuO$_4$ and La$_{2-x}$Ba$_{x}$CuO$_4$ results in quenched disordered states). This phenomenon is known as `staging' and has been well studied in the bulk compound~\cite{Wells1996,Wells1997,Blakeslee1998,Lee1999, Wochner1996, Xiong1996}. The result is a superlattice of tilt-sense antiphase boundaries along the $c$-axis, whose stage period depends on the amount of dopant oxygen $y$ (Fig.~\ref{f:f1}a,b). These planar antiphase boundaries reduce the global energy cost associated with the tilt reversals caused by the intercalated dopant oxygens. As the staging effect modulates the phase of the orthorhombic tilt pattern along the [001] direction, it can be observed in a diffraction experiment as a set of satellite peaks along the out-of-plane $L$-direction surrounding the orthorhombic tilt peaks described above. For instance, in the bulk compound a period $N$ stage order manifests peaks of the kind (0,1,$2n\pm\frac{1}{N}$), where $n$ is an integer and $N$ is measured in units of the $c$-axis lattice parameter. The films cited in this work are twinned and the measured  lattice constants are $a=5.40$~{\AA} in the plane, and $c=13.25$~{\AA}. On average, the La$_2$CuO$_{4}$ films have partially relaxed the compressive strain of LSAO.

\noindent\textbf{Acknowledgements}

We would like to thank Z. Islam, P. Wochner, M. Minola, R. Merkle, L. Balents, and D. Schlom for fruitful discussions. A.F. acknowledges support from the University of California Presidential Postdoctoral Fellowship. A.F., Z.X., M.Y., B.F., and R.J.B acknowledge financial 
support from the U.S. Department of Energy, Ofﬁce of Science, Ofﬁce of Basic Energy, Materials Sciences and Engineering Division, of the U.S. Department of Energy (DOE) under Contract No. DE-AC02-05-CH11231 within the Quantum Materials Program (KC2202). XRD experiments were performed at the Stanford Synchrotron Radiation Lightsource which is supported by the U.S. Department of Energy, Office of Science, Office of Basic Energy Sciences under Contract No. DE-AC02-76SF00515. Work at Brookhaven National Laboratory was supported by the U.S. Department of Energy, Office of Basic Energy Sciences, Early Career Award Program under Award No. 1047478. Brookhaven National Laboratory is supported by U.S. Department of Energy, Office of Science, Office of Basic Energy Sciences, under Contract No. DE-SC0012704. M.B., E.B., and B.K. acknowledge financial support from the Deutsche Forschungsgemeinschaft within the TRR80.

\noindent\textbf{Conflict of Interest}

The authors declare no conflict of interest.

\end{document}